\documentclass{ws-ijmpa}

\def\rt{\rightarrow}
\def\to{\rightarrow}

\begin{document}

\markboth{STEPHEN L. OLSEN}
{BELLE RESULTS ON CP VIOLATION}

%
\catchline{}{}{}{}{}
%

\title{RECENT BELLE RESULTS ON CP VIOLATION.}

\author{STEPHEN L. OLSEN\footnote{Permanent address: Physics
Department, University of Hawai'i, Honolulu, HI 96822 USA}}

\address{Institute of High Energy Physics 
19 Yuquan Lu, Beijing, 100049, China}

%

\maketitle


\begin{abstract}
The Belle experiment's recent results on $CP$ violation 
in $B$ meson decays are summarized.
\end{abstract}

\ccode{~~~}


In the Standard Model (SM), $CP$ violation is due to the
spontaneous breaking of electroweak symmetry 
that results in the Cabibbo-Kobayashi-Maskawa (CKM) quark 
mixing matrix.\cite{KM}
In particular, the SM predicts 
a time-dependent $CP$ violating asymmetry in the  
rates for $B^0_d$ and $\bar{B_d^0}$
decays to a common $CP$ eigenstate, $f_{CP}$: 
\setcounter{section}{1}
\begin{equation}
\label{eq:asymmetry}
A(t)\equiv 
\frac{dN/dt(\bar{B^0_t}\to f_{CP})-dN/dt({B^0}_{t}\to f_{CP})}
{dN/dt(\bar{B^0_t}\to f_{CP})+dN/dt({B^0}_{t}\to f_{CP})}
 =-\xi_f\sin 2\phi_i \sin\Delta m_d t,
\end{equation}
where $t$ is the proper time,
$dN/dt(\bar{B^0_t}$ $(B^0_{t})$ 
$\to f_{CP})$ is the decay rate
for a $\bar{B^0}(B^0)$ produced at $t=0$ to decay to $f_{CP}$ at time $t$,
$\xi_f$ is a $CP$-eigenvalue of $f_{CP}$, 
$\Delta m_d$ is the mass difference between two $B^0_d$ mass eigenstates, and
$\phi_i$ is an internal 
angle of the CKM Unitarity Triangle.\cite{Sanda}
For $CP$ eigenstate decays that proceed via a $b\rt c$ tree diagram 
({\it e.g.} $B^0\rt K_S J/\psi$), 
$\phi_i = \phi_1\equiv \arg(-(V^*_{cb}V_{cd})/(V^*_{tb}V_{td}))$,
where $V_{ij}$ ($i = u,c$ or $t$ and $j=d,s$ or $b$) are elements
of the CKM quark-flavor mixing matrix.   For $B^0 \rt K^0 J/\psi$
decays, the theoretical uncertainty
associated with the determination of $\sin 2\phi_1$ is very
small, at the 0.01 level.\cite{Beneke:2003}

\begin{figure}[tbh!]
\centerline{\psfig{file=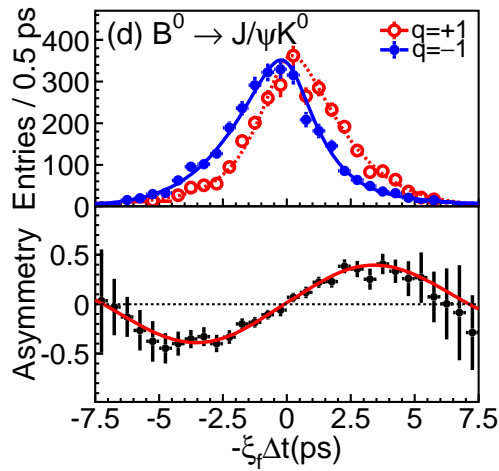,width=2.75in}}
\caption{(Top) Background-subtracted $-\xi_f\Delta t$
distributions for $B^0\rt K^0 /\psi$ events with 
high-quality  $B^0$ (open circles)
and $\bar{B^0}$ (closed circles) tags.  (Bottom) The 
$-\xi_f\Delta t$-dependent asymmetry.
\label{fig:k0jpsi}}
\end{figure}

Belle\cite{Belle_CPV2001} and BaBar\cite{BaBar_CPV2001}
first established non-zero values for $\sin 2\phi_1$ in 2001.
This year, using 7484 $B^0\rt K_S J/\psi$ ($\xi_f = -1$) decays and 
6512  $B^0\rt K_L J/\psi$ ($\xi_f = +1$) decays in a 
total sample of 535M $B\bar{B}$ meson pairs,
Belle has  reported 
$\sin2\phi_1 = +0.642\pm0.031{\rm (stat)}
\pm 0.017{\rm (syst)}$.\cite{Belle_CPV2007}
Figure~\ref{fig:k0jpsi} (top) shows the background-subtracted
$-\xi_f\Delta t$ distribution for events with a $B^0$ tag
open circles) and a $\bar{B^0}$ tag (closed circles). 
Figure~\ref{fig:k0jpsi} (bottom) shows the
$-\xi_f\Delta t$-dependent asymmetry. 

In the absence of competing penguin processes, 
for $CP$ eigenstate decays that 
proceed via a $b\rt u$ tree diagram {\it e.g.,}
($B^0\rt\pi^+\pi^-$, $\rho^+\rho^-$ and $\pi\rho$), 
$\phi_i = \phi_2\equiv \arg(-(V^*_{tb}V_{td})/(V^*_{ub}V_{ud})).$
The influence of penguins can be determined from
measurements of the branching fractions and asymmetries
of all isospin partner decays.\cite{Gronau:1990}
Because $b\rt u$ decays are suppressed ({\it e.g.} in
the data sample in which Belle finds
$\sim 14K $ $B\rt K^0 J/\psi$ decays, only
1464 $B^0\rt \pi^+\pi^-$ decays are seen) and also because of
complications from penguins, the
precision on $\phi_2$ is reduced compared to that
for $\phi_1$.  In addition, there are discrete ambiguities
associated with extracting $\phi_2$ from
the measured asymmetry.  For the 
$\pi^+\pi^-$ analysis, Belle\cite{Belle_pipi}
finds a solution consistent with SM expectations of
$\phi_2 =(97 \pm 11)^o$.  Most of the 
ambiguous solutions are ruled out by time-dependent analyses 
of $B^0\rt \pi\rho$\cite{Belle_pirho} 
and $B^0\rt\rho\rho$.\cite{Belle_rhorho}

The third angle of the Unitary Triangle, 
$\phi_3\equiv \arg(-(V^*_{ub}V_{ud})/(V^*_{cb}V_{cd})),$
is the most difficult to measure. Most proposed techniques
involve interference between the Cabibbo-suppressed but CKM-favored
$b\to c\bar{u}s$-tree-mediated $B^-\rt K^{(*)-} D^{(*)0}$ decay
with the CKM-disfavored   $b\to u\bar{c}s$-tree-mediated  
$B^-\rt K^{(*)-} \bar{D^{(*)0}}$ decay.  Here, since the 
amplitude for the latter decay is suppressed relative to the 
former both by a CKM factor  
of $|V^*_{ub}V_{cs}/V^*_{cb}V_{us}|\sim0.38$ and a color-suppression 
factor estimated to be in the range $0.1\sim 0.2$,
the expected $CP$ asymmetries are expected to be small.
The most promising technique\cite{Dalitz_bondar}   
for measuring $\phi_3$ uses
differences in the $K_S \pi^+\pi^-$ Dalitz-plot
densities between $B^+$ and $B^-$ decays to ``$D^0 " K^{\pm}$,
with ``$D^0 " \rt K_S\pi^+\pi^-$.
The gist of the technique is as follows.  Assuming no
$CP$ asymmetry in the $D^0\rt K_S \pi^+\pi^-$ decay,
the amplitude for a ``$ D^0 " \rt K_S \pi^+\pi^-$ decay
from $B\rt KD$ decays is
$M_{\pm} = f(m^2_+ ,m^2_-)+r e^{i(\delta \pm \phi_3)} f(m^2_- ,m^2_+)$,
where $m^2_{\pm} = m^2_{K_S\pi^{\pm}}$ are the two Dalitz plot
variables, $ f(m^2_+ ,m^2_-)$  is the amplitude for
$\bar{D^0}\rt K_S\pi^+\pi^-$ decay and  $r$ \& $\delta$
are the ratio and relative strong phase of the $B\rt K D$
and $B\rt K\bar{D}$ decay amplitudes.

In Belle,\cite{Belle_phi3} $ f(m^2_+ ,m^2_-)$ is parameterized
by 18 two-body decay amplitudes with
strengths and relative phases 
determined from a fit to a large
(262K event) sample of $D^+\rt \pi^+ D^0 (K_S\pi^+\pi^-)$ and charge
conjugate decays produced via the continuum $e^+e^-\rt c\bar{c}$ 
annihilation process.   The resulting  $f(m^2_+ ,m^2_-)$ 
is used in a fit of $|M_{\pm}|^2$ to the Dalitz distributions for
331 $K^{\pm}D$, 81 $K^{\pm} D^{*0}$
and 54 $K^{*\pm} D^{0}$ signal events found in a data
sample containing 386M $B\bar{B}$ meson pairs.  
Values of $r,\delta$ and $\phi_3$ for each mode are extracted 
from the fit parameters.
The value for $\phi_3$ in the range $0\le\phi_3\le 180^o$
for three combined modes is 
$\phi_3 = 53^{0}~^{+15^{0}}_{-18^{0}}{\rm (stat)}\pm 3^{0}{\rm 
(syst)}\pm 9^{0}{\rm (model)}$.
A difficulty is the sensitivity of the measured $\phi_3$ value 
to $r$, which is measured to be small and only
$\sim 3\sigma$ from zero. 
For the $K^{\pm}D$ mode the measured value
is $r=0.159 ^{+0.054}_{-0.050} \pm 0.013 \pm 0.049$.

\begin{figure}[!htb]
  \parbox{2.30in}{
        \epsfxsize=2.25in
     \centerline{\epsffile{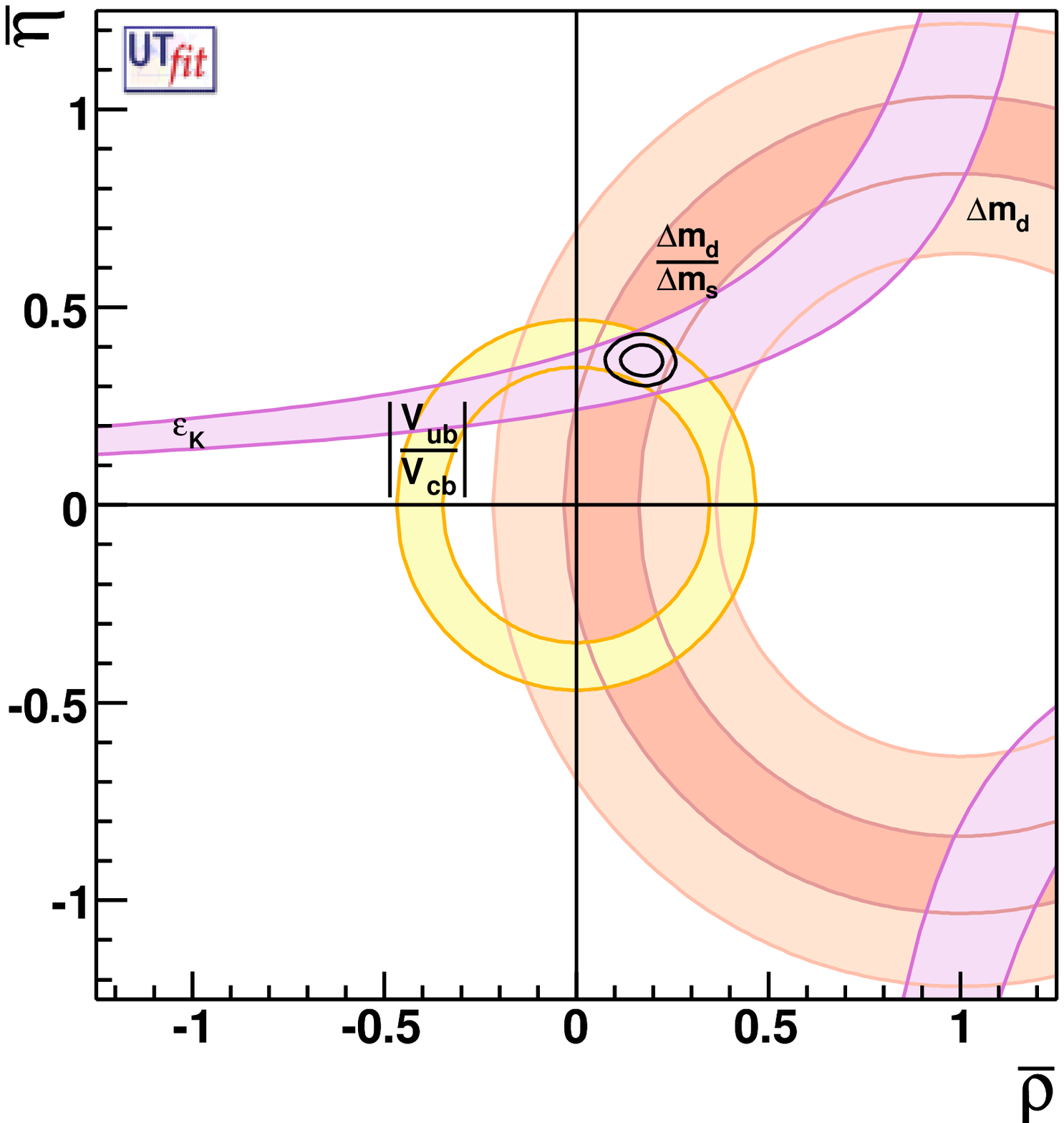}}
}
   \hspace{1mm}
  \parbox{2.30in}{
        \epsfxsize=2.25in
     \centerline{\epsffile{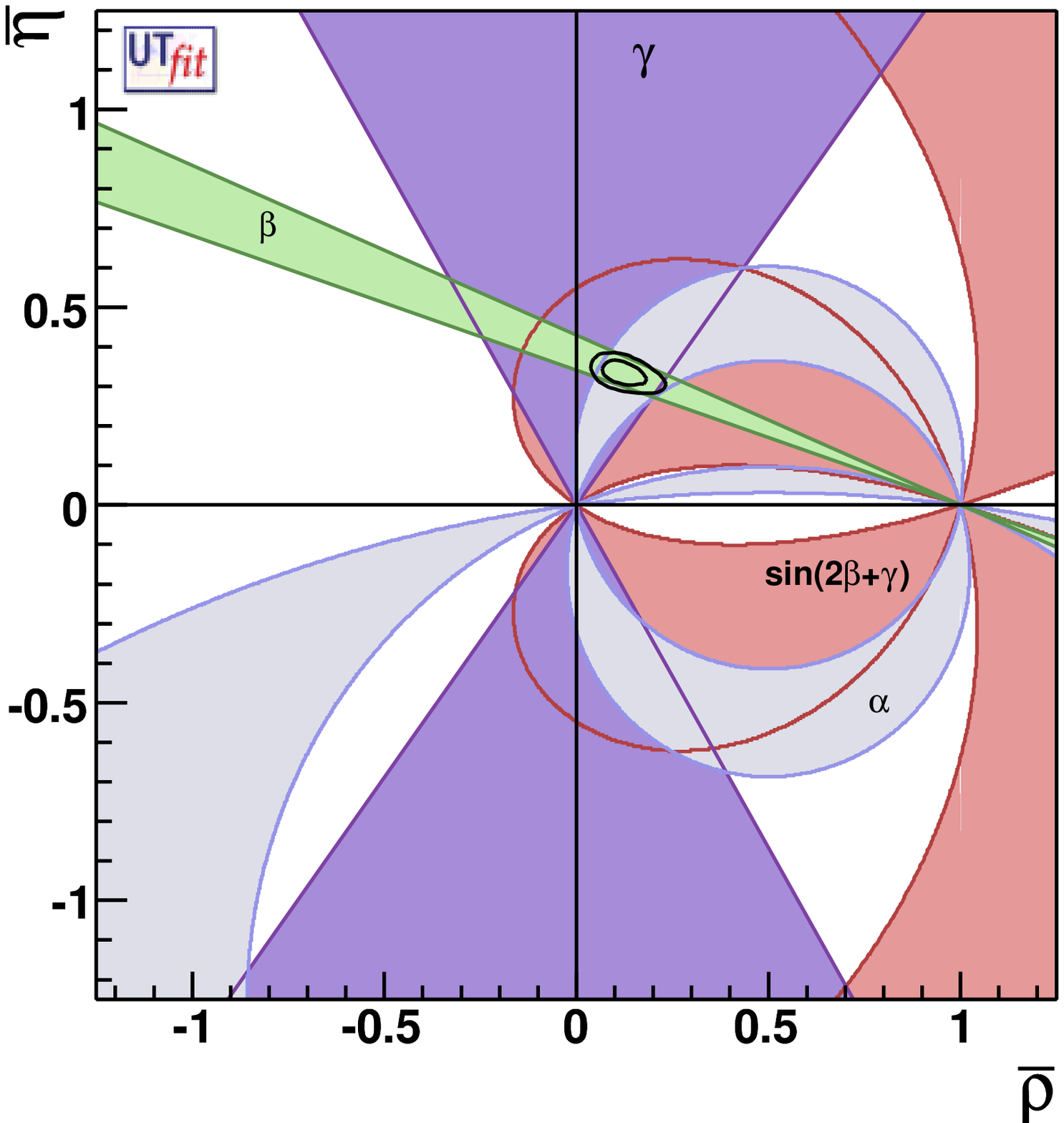}}
}
\caption{\label{fig:ckmplots}
UTfit constraints on $\bar{\rho}-\bar{\eta}$ derived
{\bf (a)} 
only from measurements other than 
$\phi_1$, $\phi_2$ or $\phi_3$;
{\bf (b)} only
from $\phi_1,\phi_2$ and $\phi_3$ measurements.
Here BaBar results are included.}
\end{figure}

The current experimental situation involving tree diagrams is
succinctly summarized in the two plots from the UTfit
group\cite{UTfit} shown in Fig.~\ref{fig:ckmplots}.
The plot on the left shows constraints in the $\bar{\rho}$-$\bar{\eta}$
plane using all measurements {\it other than} those of
the Unitary Triangle angles.  The plot on the right shows
the constraints that derive only from measurements of
$\phi_1$, $\phi_2$ and $\phi_3$.  The fact that both sets 
of measurements pick out the same allowed region in 
$\rho$ and $\eta$ is striking evidence for the validity
of the KM anzatz.  A second striking feature that is 
evident from these plots is that, in spite of the
limited precision of the $\phi_2$ measurement and
the primitive state of the $\phi_3$ measurement, the
constrains from the angle measurements are more
stringent than those from all other measurements.
Part of the reason for this is the theoretical
cleanliness of the angle measurements.  

In addition to improving the precision, especially on
$\phi_2$ and $\phi_3$, which can be done with increased
statistics, the next step is to measure
the same angles using processes involving loop (or
penguin) diagrams.  Such measurements are very sensitive 
probes for new non-SM physics.  In the SM, the loop 
processes are dominated by $W$ bosons and $t$-quarks,
two of the SM's most massive particles.
Thus, effects of non-SM particles in the loop on $CP$-violating
phases can be large even for masses of order TeV or 
higher, {\it i.e.} beyond the reach of the LHC.  
In this program, $\phi_i$ values determined
from measurements of tree processes provide precise
and theoretically robust benchmarks against which to compare
the loop process measurements.

\begin{figure}[!htb]
  \parbox{2.30in}{
        \epsfxsize=2.25in
     \centerline{\epsffile{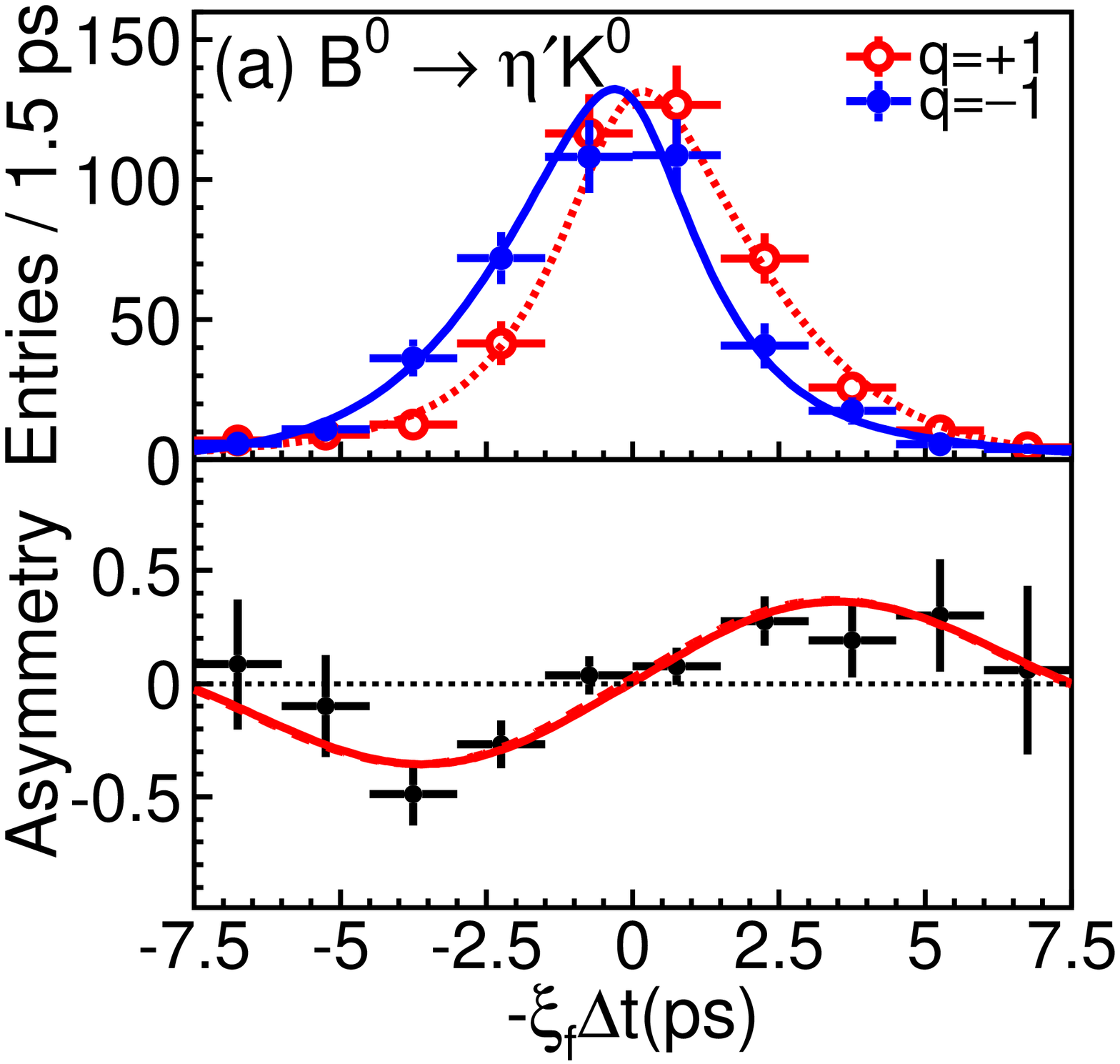}}
}
   \hspace{1mm}
  \parbox{2.30in}{
        \epsfxsize=2.25in
     \centerline{\epsffile{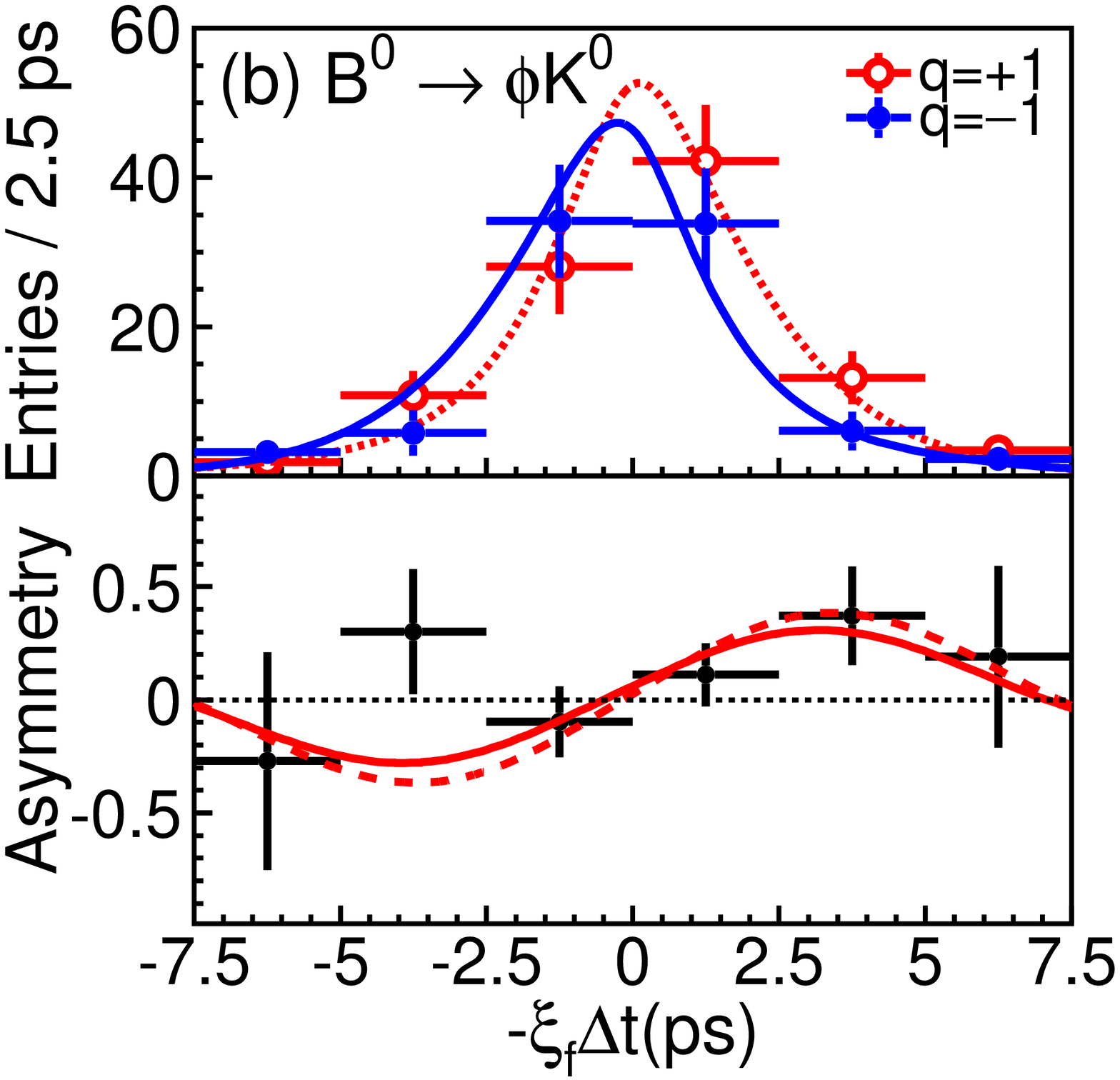}}
}
\caption{\label{fig:phiks}
$-\xi_f \Delta t$-dependent asymmetries for the
{\bf (a)} $\eta^{\prime} K^0$ and 
{\bf (b)} $\phi K^0$ decay modes.
}
\end{figure}

In the SM, measurements of mixing-induced $CP$ asymmetries
with $b\rt s$ penguin decays to $CP$ eigenstates should 
give a value of  $\sin 2\phi_1^{eff}$ that is equal to $\sin 2\phi_1$ 
to high precision . For many $b\rt s$ penguin processes, 
such as $B^0\rt\phi K^0$,
the experimental circumstances are nearly identical to those
for $B^0\rt J/\psi K^0$ and, so, comparisons can be made
with small systematic uncertainties.  The main issue is 
statistics: the branching fractions for $b\rt s$ decays are usually
about 1\% that for $J/\psi K^0$. 
Belle has made measurements of $\sin 2\phi_1^{eff}$ for an assortment
of $b\rt s $ CP eigenstate modes.\cite{Belle_CPV2007,Belle_b2s}  
The $-\xi_f\Delta t$-dependent
asymmetries for the $\eta^{\prime} K^0$ and $\phi K^0$ modes
are shown in Figs.~\ref{fig:phiks}(a) and (b), respectively;
$CP$-violating asymmetries are evident in both cases. For
these modes, $\sin 2\phi_1^{eff}$ is found to be
$+0.64\pm0.10{\rm (stat)}\pm0.04{\rm (syst)}$ ($\eta^{\prime} K^0$) and
$+0.50\pm0.21{\rm (stat)}\pm0.06{\rm (syst)}$ ($\phi K^0$). In both
cases there is good consistency with $\sin 2\phi_1$ from $B^0\rt J/\psi 
K^0$.  Measurements for other $b \rt s$ modes 
indicate no dramatic deviation
from the SM expectation but with poorer 
statistical precision.  The implication is
that any new physics that may be accessible at the
LHC has to be carefully hidden from the quark-flavor sector. 

Since SM predictions for $CP$-violating phases are free of
hadronic corrections, tests of SM predictions for these
angles are reliable and theoretically robust.  Moreover, the experimental
techniques to measure them are mature and well understood.
The primary limit is statistical precision.  Future measurements
at LHCb and, hopefully, at one or more of the proposed  
high-luminosity ``Super-B-factories,''
will provide tests of the SM at mass scales that are well beyond the
reach of the LHC.




\end{document}